\documentclass[twocolumn,aps]{revtex4}

\usepackage{graphicx}
\usepackage{dcolumn}
\usepackage{bm}
\begin{document}
\title{Resonant excitation and anti-Stokes luminescence of 
GaAs single quantum dots}
\author{K. Edamatsu, C. Watatani, T. Itoh, S. Shimomura, and S. Hiyamizu}
\address{
Division of Materials Physics, Graduate School of Engineering Science, 
Osaka University, Toyonaka 560-8531, Japan}
%
%
%
\begin{abstract}
We have investigated micro-photoluminescence ($\mu$-PL) and 
excitation ($\mu$-PLE) spectra of a single GaAs/AlGaAs quantum dot 
grown on a GaAs (411)A surface.
We observed sharp resonant lines in both $\mu$-PL and $\mu$-PLE spectra,
corresponding to the discrete energy levels of the dot. 
When the sample was excited at one of the resonant lines,
resonant luminescence lines appear not only in Stokes side but also in 
anti-Stokes side.
We discuss the possible origins of the anomalus anti-Stokes luminescence.
\end{abstract}
%
%
%
\maketitle
\section{Introduction}
Semiconductor quantum dots (QDs) or nanocrystals attract much attention
because of their unique and interesting optical properties originating 
from three-dimensional quantum confinement of the electronic states.
Single quantum dot spectroscopy has revealed many interesting and 
important properties of the QDs, 
such as sharp homogeneous linewidth \cite{Gammon}, long coherence 
time \cite{Bonadeo}, luminescence intermittency \cite{Nirmal}, 
and anti-bunched photon emission \cite{Michler}. 
In this paper, we report the micro-photoluminescence ($\mu$-PL) 
and micro-photoluminescence excitation ($\mu$-PLE) spectra from 
a GaAs/AlGaAs quantum dot grown on the (411)A surface of a GaAs substrate. 
We found anomalous anti-Stokes luminescence that is 
probably caused by efficient Auger or cascade excitation process
of the excitions in a QD.

\begin{figure}
\includegraphics[width=80mm]{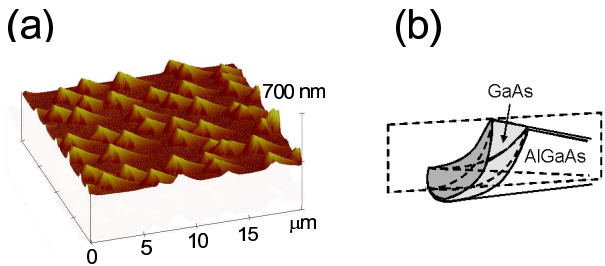}
\caption{
(a) Atomic force microcopy image of the GaAs (411)A surface of the sample. 
The scale of the image is 20 $\mu$m $\times$ 20 $\mu$m.
(b) Schematic drawing of the GaAs/AlGaAs quantum well structure overcoated on the pyramidal structure.}
\label{fig-1}
\end{figure}

\section{Experimental Procedure}
Our GaAs quantum dot sample is based on a GaAs/Al$_{0.3}$Ga$_{0.7}$As quantum well (QW) on a GaAs (411)A surface, fabricated by a molecular beam epitaxy method \cite{Hayashi}. 
Under an appropriate growth condition with relatively small As flux, 
triangular pyramidal-shaped structures are formed on the surface. 
Figure 1(a) shows the atomic force microscope (AFM) image of the pyramidal structures. 
By growing a QW over the pyramid, 
thicker area of the well is formed on a slope of the pyramid, as shown 
in Fig. 1(b). 
The mean thickness of the GaAs layer of the sample used for this experiment
was 3 nm.
The QW in the region of the pyramidal structure
contains a number of localized QD structures, 
as described later. 
Since the density of the pyramid is as low as 10$^7$cm$^{-2}$, 
we can easily resolve the luminescence from a single pyramidal 
structure using a conventional microscope objective lens. 

We used a continuous wave Ti:Sapphire laser as the excitation light source 
for the $\mu$-PL and $\mu$-PLE measurement.
The spectral width of the excitation laser was 0.17 meV.
The excitation beam was focused by an objective lens ($\times$50, N.A.=0.42)
onto the sample, which was mounted on the cold-finger of a cryostat 
and cooled down to 4 K. 
The luminescence from the sample was collected by the same objective lens.
The collected light was dispersed by a triple-grating monochromator 
(focal length = 64 cm)
and then detected by a cooled charge coupled device (CCD) camera. 
The spectral resolution of the detection system was 0.09 meV.

\begin{figure}
\includegraphics[width=65mm]{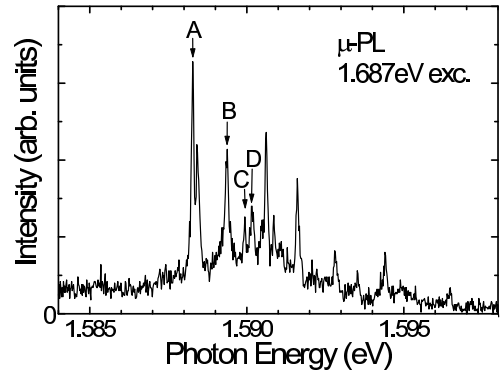}
\caption{
Micro-photoluminescence ($\mu$-PL) spectra of the GaAs QDs at 4 K.
The sample was excited at $h\nu$=1.687 eV, with power density 5.5 W/cm$^2$. }
\label{fig-2}
\end{figure}

\section{Results and Discussion}
The $\mu$-PL spectrum observed from a single pyramidal structure 
under moderately weak excitation power density (5.5 W/cm$^2$)
is shown in Fig. 2.
The excitation was made at 1.687 eV,
which corresponds to the excitation above the GaAs QW 
surrounding the pyramidal structure.
Several sharp luminescence lines were observed in the $\mu$-PL spectrum.
However, from the PL spectrum, it is difficult to determine the number of QDs 
and their energy levels corresponding to these lines.
In order to observe this, 
we measured the $\mu$-PLE spectra detected at four luminescence lines (A to D in Fig. 2), as shown in the upper graph of Fig. 3.
For comparison,
the $\mu$-PL spectrum identical to Fig. 2 is also shown in the lower 
graph of Fig. 3 with the same energy scale as the $\mu$-PLE spectra, 
The $\mu$-PLE spectra have several sharp peaks reflecting the discrete energy levels of the QD. 
One can see the close similarity between the PLE spectra for A and D, 
and for B and C.
The $\mu$-PLE spectrum for A has a sharp peak at the energy posision
of the PL line D.
Also, the $\mu$-PLE for B has a sharp peak at the PL line C.
These findings suggest that the luminescence lines A and D originate 
from an identical dot, and lines B and C from another dot in a pyramid. 
Thus, we conclude that there are a number of (at least two) quantum dots
in the single pyramidal structure.
In addition, for all the PLE spectra, 
quasi-continuous excitation band appears above 1.594 eV. 
The continuous PLE band suggests that the quantum dots are surrounded by the two-dimensional states of the adjacent GaAs layer. 

\begin{figure}
\includegraphics[width=75mm]{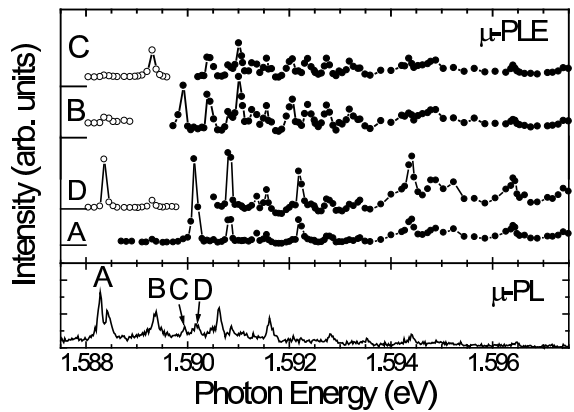}
\caption{
Micro-photoluminescence excitation ($\mu$-PLE: upper) and
micro-photo\-luminescence ($\mu$-PL: lower) spectra of the GaAs QDs at 4 K. 
The $\mu$-PLE spectra A$\sim$D were detected at the four sharp luminescence
lines A$\sim$D indicated in the $\mu$-PL spectrum, respectively. 
Excitation intensities for the $\mu$-PLE spectra were 55 W/cm$^2$ 
and 110 W/cm$^2$ for the Stokes (filled circles) and 
anti-Stokes (open circles) spectra, respectively. }
\label{fig-3}
\end{figure}

\begin{figure}
\includegraphics[width=65mm]{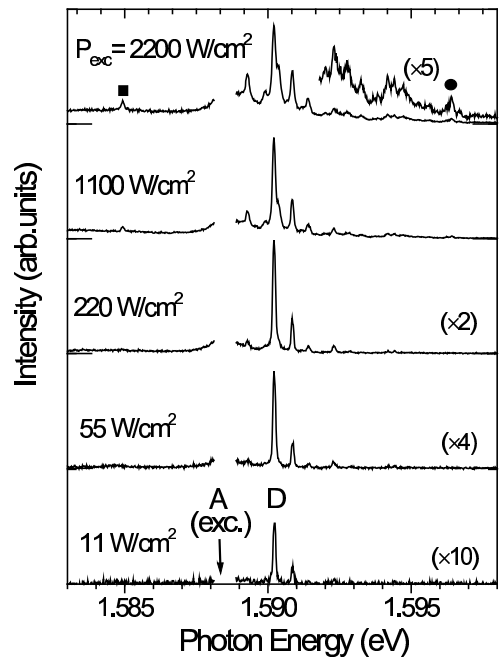}
\caption{
Resonant micro-photoluminescence spectra of the GaAs QDs at 4 K
at various excitation intensities $P_{\rm exc}$. 
The excitation was made at the energy position ($h\nu$=1.5882 eV) 
indicated by the downward arrow, 
which corresponds to the luminescence line A indicated in Figs. 2 and 3.}
\label{fig-4}
\end{figure}

Furthermore, we found that
the $\mu$-PLE spectra for the PL lines D and C have sharp anti-Stokes 
resonances at the energy positions of the PL lines A and B, respectively,
as indicated by the open circles in Fig. 3.
The anti-Stokes shifts are 1.9 meV and 0.6 meV for the two cases, respectively.
Since the spectra were taken at low temperature (4 K), the anti-Stokes PL cannot be caused by any thermal excitation. 
Local heating caused by the excitation light was not expected
because the spectral position of the luminescence lines did not change
noticeably in the range of the excitation intensity (see Fig. 4),
although the expected spectral red shift would have been $\sim$0.2 meV 
if the sample temperature had been raised to 20 K ($kT$=1.7 meV).
To our knowledge, such distinct and anomalous anti-Stokes resonant 
luminescence has not been reported for any kind of semiconductor QDs 
or nanocrystals.
Figure 4 shows the $\mu$-PL spectra for various excitation intensities 
($P_{\rm exc}$)
under resonant excitation at the energy position corresponding to the PL line A. 
One can see that several PL lines appear on the anti-Stokes sides, 
including the distinct line at D, which corresponds to the PL line D in
Figs. 2 and 3.
The energy shift of the anti-Stokes PL 
from the excitation photon energy ranges more than 8 meV.
From the analysis of the excitation-intensity dependence, 
we found that the intensity of the anti-Stokes PL line at D 
has almost linear dependence on the excitation intensity
under weak or moderate excitation ($P_{\rm exc}$$\sim$100 W/cm$^2$ or less).
Since the thermal excitation is not expected as mentioned above, 
the origin of the linear anti-Stokes PL is an open question at present.
In contrast, 
the intensity of PL lines with larger anti-Stokes shift
has considerable nonlinear dependence on the excitation intensity.
For instance, 
the intensity of the anti-Stokes PL line indicated by the filled circle 
in Fig. 4 was found to have
quadratic dependence on the excitation intensity.
This quadratic excitation-intensity dependence indicates that 
the anti-Stokes PL line arises when two excitons are created, 
or two photons are annihilated, in the QD.
One possible origin of the anti-Stokes PL is the
Auger process between the two excitons in the QD,
which often causes luminescence intermittency \cite{Nirmal,Efros}.
Another possibility is the cascade two-step excitation of the exciton in 
the QD.
Both processes excite the exciton to delocalized or ionized states,
from which the electon-hole pair can relax into 
the initial state of the anti-Stokes luminescence.
Furthermore, under high excitation intensity,
another PL line (filled square in Fig. 4) appears in the lower energy side. 
Its intensity also has quadratic dependence on the excitation intensity.
This line is attributable to the emission from the biexciton in the QD. 
The binding energy of the biexciton is 3.4 meV in this case, 
which is comparable to the  reported experimental \cite{Brunner} 
and theoretical \cite{Tsuchiya}
values for GaAs QDs. 
Thus, it is likely that the two-exciton bound state (biexciton)
coexists with the Auger or two-step excitation process of the excitons
in the QD.
The coexistence of the two contradictory processes
is an interesting and important problem
in discussing the few-particle dynamics in the QD.
Further investigation is in progress
in order to identify the origin of the anomalous resonant 
anti-Stokes luminescence.

\section*{Acknowledgements}
This work was supported by Grant-in-Aid for COE Research (10CE2004) and Grant-in-Aid for Scientific Research from the Ministry of Education, 
Culture, Sports, Science and Technology of Japan.

\newpage

\end{document}